\documentclass[prb,twocolumn]{revtex4-1}

\usepackage{amsmath}  
\usepackage{graphicx} 
\usepackage[utf8]{inputenc}
\usepackage{bm,color}

\newcommand{\abs}[1]{\left\lvert{#1}\right\lvert}

\begin{document}

\title[Momentum and energy in special relativity]{Elementary derivation of the expressions of momentum and energy in special relativity}

\author{Luca Peliti}
\email{luca@peliti.org}
\affiliation{M. A. and~H. Chooljan Member, Simons Center for Systems Biology, Institute for Advanced Study, Einstein Drive, Princeton NJ 08540 (USA)}

\date{February 6, 2016}

\begin{abstract}
The derivation of the expressions of momentum and energy of a particle in special relativity is often less than satisfactory in elementary texts. In some, it is obtained by resorting to quantum or electrodynamic considerations, in others by introducing less-than-elementary concepts, like that of a four-vector, or even misleading ones, like ``relativistic mass''. Nevertheless it is possible, following ideas described by Einstein in 1935, to obtain a fully elementary derivation of these expressions based only on the Lorentz transformations, on the conservation laws, and on the Newtonian limit. The resulting argument allows for a clearer and logically consistent introduction to the basic concepts of relativistic dynamics.
\end{abstract}

\maketitle 

\section{Introduction} 
\label{sec:intro}
Several texts provide an elementary derivation of the kinematics of special relativity, eventually based on the first part of Einstein's fundamental paper of 1905.~\cite{Einst05} Starting from the two postulates of the total equivalence of inertial reference systems, and of the constancy of the speed of light in all such systems, it is in fact easy to obtain the expression of the Lorentz transformation of space-time coordinates. A particularly simple and appealing derivation is obtained by exploiting Bondi's so-called $k$-calculus,~\cite{Bondi80} itself based on the Doppler effect. Nevertheless going from kinematics (the Lorentz transformation) to dynamics, and in particular to the relativistic expressions of momentum and energy, is often achieved by resorting to more sophisticated concepts, like that of a four-vector, or by the use of quantum considerations. (One example of this approach is the ``elementary derivation'' suggested by F.~Rohrlich,~\cite{Rohrlich} which exploits the expressions of the momentum and energy of a photon of frequency~$\nu$.) Einstein himself had originally derived the mass-energy equivalence via an explicit use of electrodynamics,~\cite{Einst05a} and not just by the kind of kinematic considerations which he had developed in the first part of his 1905 paper, which are based on the constancy of the speed of light, but do not otherwise depend on Maxwell's equations.

Einstein had remarked this problem, and he proposed in 1935 an elementary derivation of the mass-energy relation, independent of his 1905 argument, motivating it with the following words:~\cite{Einst35}
\begin{quotation}
The special theory of relativity grew out of the Maxwell
electromagnetic equations. So it came about that even in the
derivation of the mechanical concepts and their relations the
consideration of those of the electromagnetic field has played
an essential role. The question as to the independence of those
relations is a natural one because the Lorentz transformation,
the real basis of the special relativity theory, in itself has nothing
to do with the Maxwell theory and because we do not know the
extent to which the energy concepts of the Maxwell theory can
be maintained in the face of the data of molecular physics. In
the following considerations, except for the Lorentz transformation,
we will depend only on the assumption of the conservation
principles for impulse and energy. 
\end{quotation}
Einstein's considerations exploit a conceptual experiment introduced by G.N.~Lewis and R.C.~Tolman~\cite{Lewis09} and further discussed by P.S.~Epstein,~\cite{Epstein11} where one considers collisions between pairs of particles in different inertial reference frames, and one looks for the expressions of momentum and energy by postulating their conservation. It is interesting to point out that Lewis and Tolman's paper, as well as Epstein's one, only consider elastic collision and derive the relativistic expression of momentum, while they provide a doubtful argument for the mass-energy equivalence by the consideration of the change of the ``relativistic mass'' with speed. ``Relativistic mass'' is in fact a rather problematic concept,\cite{Adler,Okun} and there is a growing consensus to avoid its introduction in the teaching of relativity. Einstein derives instead the equivalence by simply extending the argument to inelastic collisions. The advantage of this approach for introducing the basic concepts of special relativity has been well remarked by R.F.~Feynman who, in his \textit{Lectures}~(Ref.~\onlinecite[Vol.~I. Secs.~16--4,~16--5]{Feynman}), derives the relativistic expressions of momentum and energy in a way that closely resembles Einstein's one. (Feynman's derivation is however marred by his use of the ``relativistic mass''.) Einstein's argument has been more recently discussed by~F.~Flores,~\cite{Flores} who identifies three closely related but different claims within the mass-energy equivalence concept, and compares Einstein's 1935 argument with his original 1905 derivation~\cite{Einst05a} and with M.~Friedman's 1983 derivation,~\cite[p.~142ff]{Friedman} which rests upon the consideration of Newton's equations in special relativity.

In this note, I present this line of thought in the hope that it may be found useful for the presentation of these fundamental concepts of special relativity in introductory courses for students of physics and mathematics. While the derivation of the relativistic expression of momentum~in Sec.~\ref{sec:qdm} is close to Epstein's and~Feynman's arguments, the discussion of the expression of the kinetic energy and of the mass-energy equivalence is closer to Einstein's one.
\section{Lorentz transformations and dynamic postulates}
\label{sec:Lorentz}
Following Einstein's~1905 paper~(Ref.~\onlinecite[\S~2]{Einst05}), the kinematic concepts of special relativity rest on the following postulates:
\begin{enumerate}
\item[(i)] The laws that govern the transformations of the state of physical systems take the same form in reference frames animated by uniform translational motion one with respect to the other.
\item[(ii)] There is a class of such reference frames (the \textit{inertial frames}) in which the speed of light assumes the same value $c$, independently of the state of motion of its source.
\end{enumerate}
We shall choose from now on units in which $c=1$ and limit our considerations to inertial frames. Based on these postulates one can easily derive Lorentz transformations in the following form. Let us consider two reference frames, $K$ and $K'$, such that $K'$ is in uniform translational motion in the positive~$x$ direction and with speed~$V$ with respect to~$K$. Then the event of coordinates $(t',x',y',z')$ in $K'$ has in $K$ the coordinates $(t,x,y,z)$, where
\begin{equation}\begin{split}
t&=\gamma(V)(t'+V x');\\
x&=\gamma(V)(x'+V t');\\
y&= y'; \\
z&= z',
\end{split}\end{equation}
and we have defined
\begin{equation}
\gamma(V)=\frac{1}{(1-V^{2})^{1/2}}.
\end{equation}
The same relation holds for the differentials~$d  t$, $d  x$, etc. Dividing by $d  t$ we obtain the rules for the transformation of velocities:
\begin{equation}\begin{split}\label{eq:velocities}
u_{x}&=\frac{d  x}{d  t}=\frac{u'_{x}+V}{1+u'_{x}V};\\
u_{y}&=\frac{d  y}{d  t}=\frac{u'_{y}}{\gamma(V)(1+u'_{x}V)}; \\
u_{z}&=\frac{d  z}{d  t}=\frac{u'_{z}}{\gamma(V)(1+u'_{x}V)}.
\end{split}\end{equation}

To introduce dynamical concepts we obviously need supplementary postulates. We shall therefore postulate the following:
\begin{enumerate}
\item[(iii)] The momentum $\bm{P}$ and the energy $E$ of a particle possessing the velocity $\bm{u}$ in the reference frame $K$ have respectively the expressions
\begin{equation}
\bm{P}=m \bm{u}\,F(u);\quad E=E_{0}+m\, G(u),
\end{equation}
where $E_{0}$ is a constant, that can be called the \textit{rest energy}, $m$ is a positive constant (which is \textit{relativistically invariant}, i.e., does not change from one inertial frame to another, and thus \textit{does not depend on the particle's velocity}) which we shall simply call its \textit{mass}, and $F(u)$ and~$G(u)$ are \textit{monotonically increasing} universal functions of $u=\abs{\bm{u}}$. The fact that the functions $F$ and~$G$ depend only on the magnitude $u$ of the velocity, but not on its direction, can be inferred by the isotropy of space.
\item[(iv)] For $u\ll 1$ these expressions reduce to the well-known classical ones. One has in particular
\begin{equation}
F(u)=1+\mathrm{O}(u^{2});\qquad G(u)=\frac{1}{2}u^{2}+\mathrm{O}\left(u^{4}\right).
\end{equation}
\item[(v)] The total momentum $\bm{P}^{\mathrm{tot}}$ and the total energy $E^{\mathrm{tot}}$ of a system of several particles are respectively given by the sum of~$\bm{P}$ and of~$E$ running over all the particles of the system. 
\item[(vi)] \textbf{Conservation of momentum and energy:} Let us assume that (elastic or inelastic) collisions occur in a system of particles. Then, in each reference frame, $\bm{P}^{\mathrm{tot}}$ and~$E^{\mathrm{tot}}$ maintain the same values before and after each collision.
\end{enumerate}
As a corollary, the velocity of each particle remains constant as long as no collision occurs, because one can consider systems made up of single, independent particles.
\section{Elastic collisions and relativistic momentum}
\label{sec:qdm}
Let us now consider a \textit{particle pair}, i.e., a system made of two particles with equal values of~$m$. Let us assume that in a reference frame~$K$ they have opposite velocities $\bm{u}_{1}$, $\bm{u}_{2}=-\bm{u}_{1}$, where $\bm{u}_{1}=(V, v, 0)$, with $\abs{v}\ll V$, $V>0$. Thus $\abs{\bm{u}_{1,2}}\simeq V$. Let us moreover assume that the particles undergo an elastic collision, and take on respectively the velocities $\bm{w_{1}}=(W,w,0)$ and~$\bm{w_{2}}=(W',w',0)$ after it. By the conservation of momentum one must have $W'=-W$ and~$w'=-w$, \textit{independently of the form of the function~$F(u)$.} Indeed, one has $\bm{P}^{\mathrm{tot}}_{\mathrm{in}}=0$ before the collision. By the conservation law one must have
\begin{equation}
\bm{P}^{\mathrm{tot}}_{\mathrm{out}}=0=m\bm{w}_{1}\,F(w_{1})+m\bm{w}_{2}F(w_{2}).
\end{equation}
Thus the vectors $\bm{w}_{1}$ and $\bm{w}_{2}$ are antiparallel, and one has
\begin{equation}
\frac{\abs{\bm{w_{1}}}}{\abs{\bm{w}_{2}}}=\frac{F(w_{2})}{F(w_{1})}.
\end{equation}
Since the function~$F(u)$ is monotonically increasing, this equation can only be satisfied if~$\abs{\bm{w}_{1}}=\abs{\bm{w}_{2}}$, and we have therefore $\bm{w}_{1}=-\bm{w}_{2}$. Now energy conservation imposes $w_{\pm}=u_{\pm}$. Indeed, the total kinetic energy before the collision is given by $2mG(u)$, and after it is given by~$2mG(w)$. Since $G(u)$ is a monotonically increasing function of~$u$, this condition can only be satisfied if~$w_{\pm}=u$. 

\begin{figure}[htb]
\begin{center}
\includegraphics[width=0.45\textwidth]{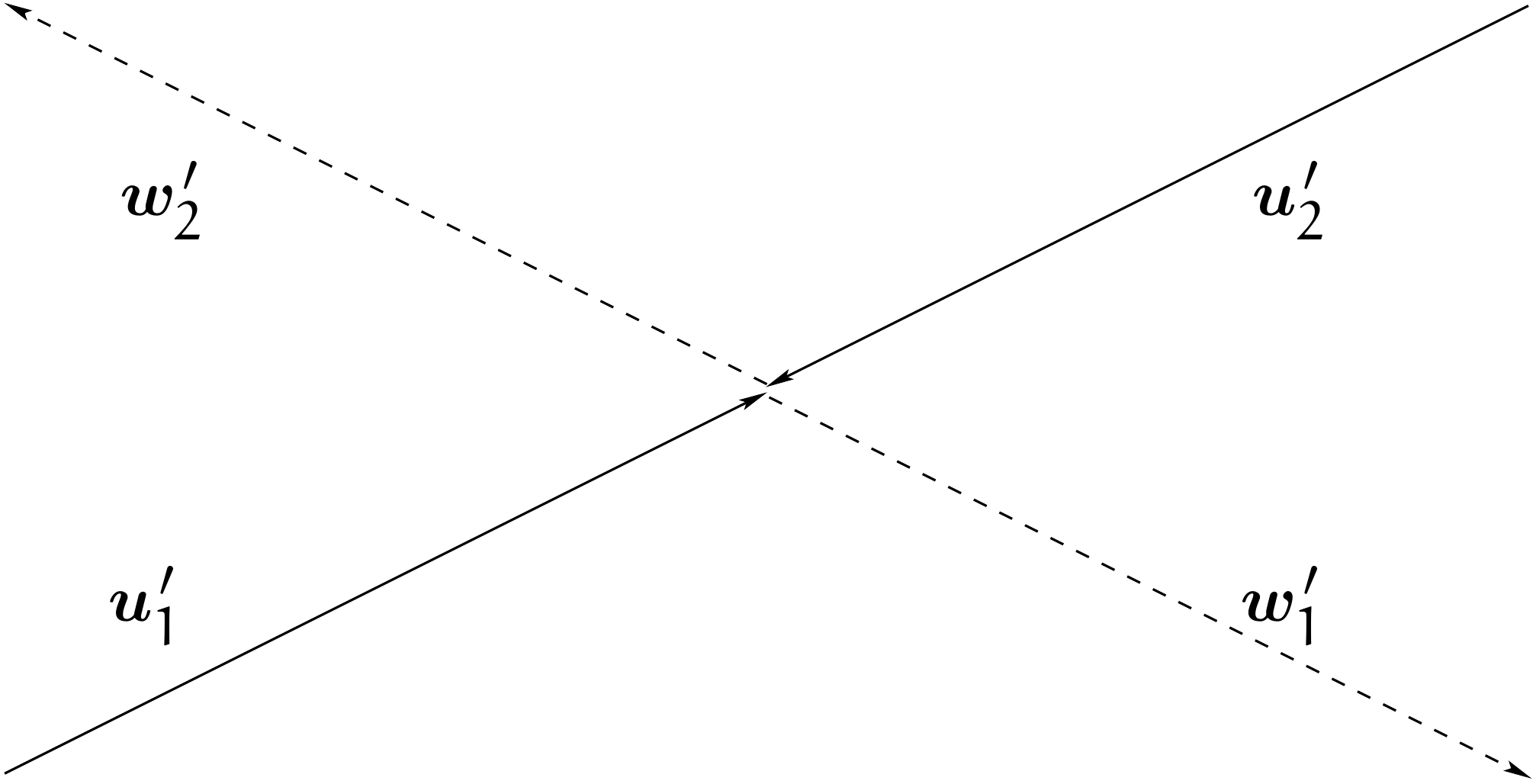}\\[0.3cm]\includegraphics[width=0.45\textwidth]{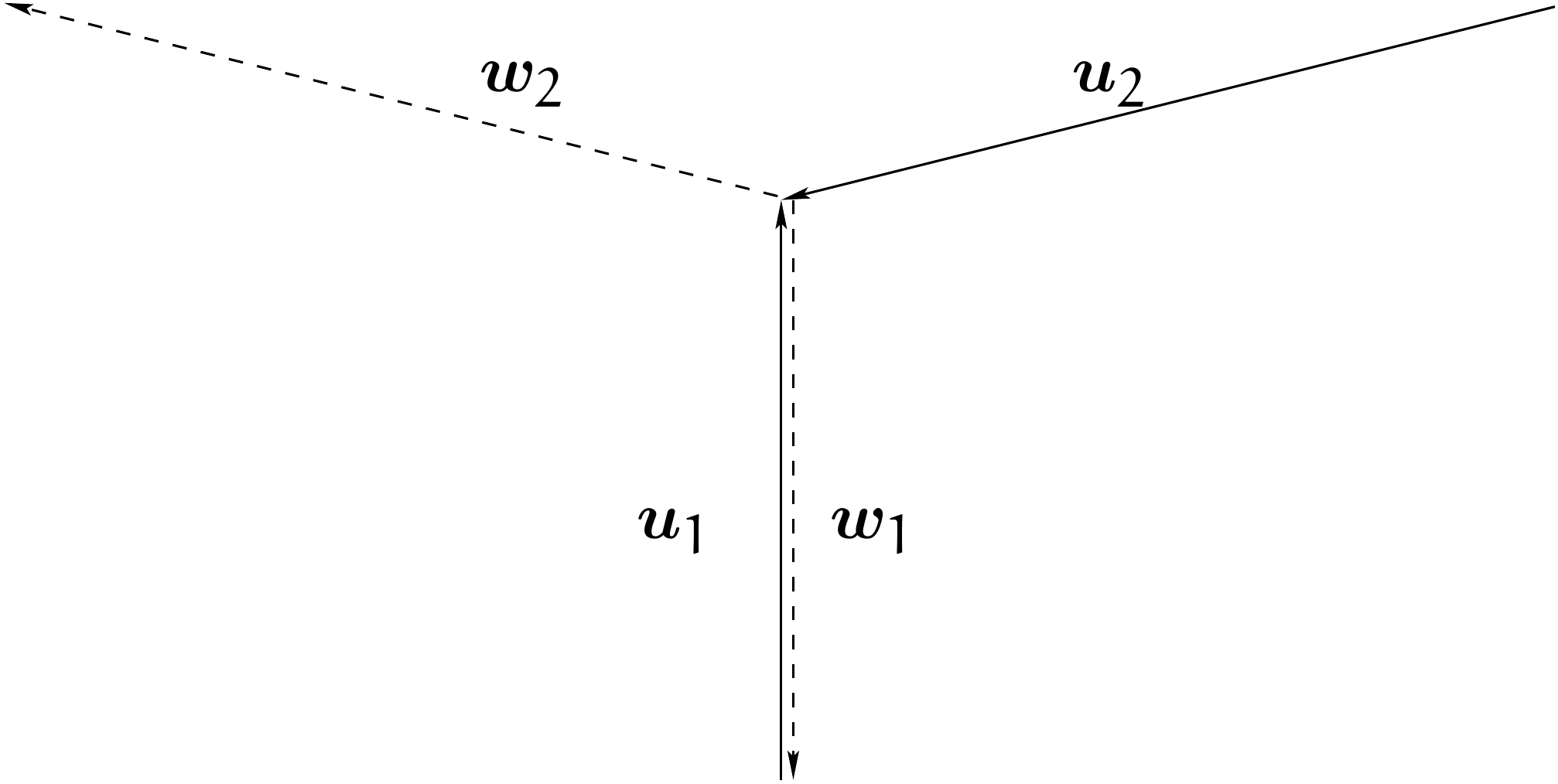}
\end{center}
\caption{Above: Collision of a particle pair in the~$K'$ reference frame. Below: The same collision as seen in the~$K$ reference frame. Figures are schematic and no quantitative relations are implied.}\label{fig:coll}
\end{figure}

Let us now consider the special case where the velocity change is parallel to the~$y$ axis. (This rules out ``head-on collisions'', where the particles simply exchange their velocity.) Let the particles' velocities be given by~$\bm{u}'_{1}=(V',v',0)$ and $\bm{u}'_{2}=-\bm{u}'_{1}$ (cf.~fig.~\ref{fig:coll}) in the $K'$ reference frame. Let us now look at the same collision in a reference frame~$K$ moving with a velocity $V'$ in the $x$ direction with respect to~$K'$.
In this reference frame, the velocities~$\bm{u}_{\pm}$ of the particles are respectively given by
\begin{equation}
\begin{split}
\bm{u}_{1}&=(0,v,0),\\
\bm{u}_{2}&=(-V,-w,0),
\end{split}
\end{equation}
where
\begin{equation}
\begin{split}
V&=\frac{2V'}{1+{V'}^{2}},\\
v&=\frac{v'}{\gamma(V')(1-{V'}^{2})},\\
w&=\frac{v'}{\gamma(V')(1+{V'}^{2})}.
\end{split}
\end{equation}

One can easily check that
\begin{equation}
\gamma(V)=\frac{1}{\sqrt{1-{V}^{2}}}=\frac{1+{V'}^{2}}{1-{V'}^{2}},
\end{equation}
and therefore that
\begin{equation}
v=w\,\gamma(V).
\end{equation}
One can also obtain this result by applying Eq.~(\ref{eq:velocities}) to the transformation from the $K'$ reference frame to a frame $K''$ animated, with respect to $K'$, by a uniform translational motion with velocity $-V'$ parallel to the~$x$ axis. In this frame the $x$ component of the velocity~$\bm{u}''_{2}$ vanishes and that of the~$\bm{u''}_{1}$ velocity is equal to $V$, and thus the speeds of the two particles are interchanged.

Let us now consider the conservation of momentum. The change $\delta \bm{P}_{1}$ of the momentum of particle~1 is given by
\begin{equation}
\delta \bm{P}_{1}=-2 m v\,F(v) \,\bm{e}_{y},
\end{equation}
where $\bm{e}_{y}$ is the~$y$ axis versor. The corresponding quantity for particle~2 is given by
\begin{equation}
\delta \bm{P}_{2}=2 m w\,F(u_{2}) \,\bm{e}_{y},
\end{equation}
where $u_{2}=\sqrt{{V}^{2}+{w}^{2}}$. Let us momentarily assume $v,w\ll V$: then $F(v)\simeq 1$ and $u_{2}\simeq V$. From momentum conservation we obtain $\delta\bm{P}_{1}+\delta\bm{P}_{2}=0$, which implies
\begin{equation}
mv=mw\,F(V).
\end{equation}
Since $w=v/\gamma(V)$, we obtain
\begin{equation}
F(V)=\gamma(V)=\frac{1}{\sqrt{1-{V}^{2}}}.
\end{equation}
Having obtained this result for $v,w\ll V$, it is easy to see that it also holds for larger values of $v$ and~$w$, by substituting $V$ with the speed of the corresponding particle. We have in fact
\begin{equation}
mv \,F(v)=m w\,F(u_{2})=\frac{mv}{\gamma(V)} \,F(u_{2}),
\end{equation}
from which it follows
\begin{equation}
F(u_{2})=F(v)\gamma(V),
\end{equation}
namely
\begin{equation}
\frac{1}{\sqrt{1-{u_{2}}^{2}}}=\frac{1}{\sqrt{1-{v}^{2}}}\frac{1}{\sqrt{1-{V}^{2}}},
\end{equation}
an identity which is easy to check directly.
\section{Kinetic energy conservation}
\label{sec:kinetic}
Let us consider a particle having the velocity $\bm{u}'=(u',0,0)$, parallel to the~$x$ axis, in the~$K'$ reference frame. Its velocity~$\bm{u}$ in the~$K$ frame, in uniform translational motion with respect to~$K'$ with speed~$V$ in the direction of the positive $x$-axis, is given by $\bm{u}=(u,0,0)$, with
\begin{equation}
u=\frac{u'+V}{1+u'V}.
\end{equation}
One can easily see that
\begin{equation}
\gamma(u)=(1+u'V)\,\gamma(u')\gamma(V).
\end{equation}
If $\bm{u}'$ is not parallel to the~$x$ axis, but one has instead $\bm{u}'=(u'_{x},u'_{y},u'_{z})$, one has the more general relation
\begin{equation}
\gamma(\abs{\bm{u}})=(1+u'_{x}V)\gamma(u')\gamma(V),
\end{equation}
which can be obtained with a little algebra. It is also easy to check that
\begin{equation}\begin{split}\label{eq:momentum}
u_{x}\gamma(u)&=(u'_{x}+V)\gamma(u')\gamma(V);\\
u_{y}\gamma(u)&=u'_{y}\gamma(u');\\
u_{z}\gamma(u)&=u'_{z}\gamma(u').
\end{split}\end{equation}

Let us now consider a particle pair, which have opposite velocities $\bm{u}'_{1}$, $\bm{u}'_{2}=-\bm{u}'_{1}$ in the $K'$ frame. Thus $u'=\abs{\bm{u}'_{1}}=\abs{\bm{u}'_{2}}$ is the common value of the particles' speed in the $K'$ frame. Let us denote by $\bm{u}_{1}$ and $\bm{u}_{2}$ the corresponding velocities in the~$K$ frame. We obtain
\begin{equation}
\gamma(u_{1})+\gamma(u_{2})=2\gamma(u')\gamma(V).
\end{equation}
We have seen that an elastic collision in the~$K$ frame cannot change the common value $u'$ of the particles' speed. Thus the right-hand side of this equation cannot change in the collision. But then neither can its left-hand side. If we denote by $\bm{w}_{1}$ and~$\bm{w}_{2}$ the particles' speeds in the~$K$ frame after the collision, we obtain
\begin{equation}\label{eq:consGamma}
\gamma(u_{1})+\gamma(u_{2})=\gamma(w_{1})+\gamma(w_{2}).
\end{equation}
As Einstein~(Ref.~\onlinecite[p.~227]{Einst35}) points out, these equations have the form of conservation laws. We can thus interpret\cite{note} $m\,(\gamma(u)-1)$ as the kinetic energy $T$ of a particle with mass~$m$ animated by a velocity~$\bm{u}$. For small values of~$u$ this quantity is thus given by
\begin{equation}
T=m\left(\gamma(u)-1\right)\simeq \frac{1}{2}mu^{2},
\end{equation}
in agreement with the classical limit. We can thus set
\begin{equation}
G(u)=\gamma(u)-1.
\end{equation}
Let us remark moreover that, by applying Eqs.~(\ref{eq:momentum}) to a particle pair, we obtain
\begin{equation}
\bm{u}_{1}\gamma(u_{1})+\bm{u}_{2}\gamma(u_{2})=2\bm{V}\gamma(u')\gamma(V).
\end{equation}
We can thus derive the following relation:
\begin{equation}
\bm{u}_{1}\gamma(u_{1})+\bm{u}_{2}\gamma(u_{2})=\bm{w}_{1}\gamma(w_{1})+\bm{w}_{2}\gamma(w_{2}),
\end{equation}
which can be interpreted as the conservation law for the momentum. We thus recover the relativistic expression of the momentum derived in sec.~\ref{sec:qdm}.
\section{Mass-energy equivalence}
\label{sec:equivalence}
Let us now consider a totally inelastic collision in a particle pair. In the reference frame $K'$ in which $\bm{P}^{\mathrm{tot}}=0$, the total kinetic energy before the collision is given by
\begin{equation}
T'_{\mathrm{in}}=2m\left(\gamma(u')-1\right),
\end{equation}
For simplicity, we shall assume that the particles velocities $\bm{u}'_{1,2}$ are parallel to the $y$ axis. The total kinetic energy after the collision vanishes, but the energy of the resulting particle has increased by~$T'_{\mathrm{in}}$. By momentum conservation the resulting particle is at rest in the reference frame~$K'$ and has thus velocity~$\bm{V}$ parallel to the $x$ axis in the~$K$ frame. Let us denote by~$M$ its mass. In the~$K$ frame the total momentum before the collision is given by
\begin{equation}
\bm{P}=m\left(\bm{u}_{1}\gamma(u_{1})+\bm{u}_{2}\gamma(u_{2})\right)=2m\bm{V}\gamma(u')\gamma(V),
\end{equation}
while after the collision it has the value
\begin{equation}
\bm{P}=M\bm{V}\gamma(V).
\end{equation}
We obtain therefore
\begin{equation}
M=2m\gamma(u')=2m+T'_{\mathrm{in}}.
\end{equation}
Therefore when two particles collide inelastically to form a new particle, the mass of the resulting particle is larger than the sum of the masses of the colliding particles by a quantity exactly equal, in our units, to the kinetic energy transformed by the collision into other energy forms. One can thus introduce a ``natural'' choice of the energy at rest $E_{0}$ of a particle, by equating it (in our units) to its mass, also because, ``from the nature of the concept, [that] is determined only to within an additive constant, one can stipulate that $E_{0}$ should vanish together with $m$.''\cite[p.~229]{Einst35}. We can then identify $m\gamma(u)$ with the total energy of a particle, and associate the change~$\delta E$ of energy from the kinetic to a different form with a change $\delta m=\delta E$ of the mass of the particle. In the more general case, in which the two particles do not coalesce, but are animated by velocities with the same magnitude $w'$ in the $K'$ frame after the collision, one can then express the conservation of energy by
\begin{equation}\label{eq:2mGamma}
2 m \gamma(u')=2\bar{m}\gamma(w'),
\end{equation}
where $\bar{m}$ is the mass of each particle after the collision. 
The same applies in the~$K$ frame, if we take into consideration the relation
\begin{equation}
\gamma(u)=\gamma(u')\gamma(V),
\end{equation}
which holds since both $\bm{u}_{1}$ and~$\bm{u}_{2}$ are parallel to the~$y$ axis.

It is then a simple matter to verify that the energy $E$ and the momentum $\bm{P}$ of a particle of mass~$m$ satisfy the relation
\begin{equation}\label{eq:EPm}
E^{2}-P^{2}=m^{2},
\end{equation}
where the right-hand side is relativistically invariant. This relation also holds in the limit of a zero-mass particle, for which it assumes the form
\begin{equation}\label{eq:EP}
E=\abs{\bm{P}}.
\end{equation}
As a corollary, \textit{mass is not additive}: the mass of a system of particles depends on its total energy content in the vanishing-momentum frame. Thus when light waves (of vanishing mass) with opposite momenta are emitted from a particle at rest, the mass of the particle changes (as argued in Einstein's 1905 work,~Ref.~\onlinecite{Einst05a}). Since the right-hand side of Eq.~(\ref{eq:EPm}) is invariant, it can be taken as the starting point to show that the energy $E$ and the momentum $\bm{P}$ combine into a relativistic four-vector, a property which generalizes to particle systems.

One should point out that this line of thought does not rest on Maxwell's equations (keeping only the constancy of the speed of light) and neither does on other mechanical concepts, in particular on the concept of force, that is difficult to justify in special relativity.  Einstein criticizes the use of the concept of force in the derivation of the relativistic expression of momentum, contained in the book by G.D. Birkhoff and~R.E.~Langer, \textit{Relativity and Modern Physics},~\cite{Birkhoff} exactly for this reason, as made clear by the closing paragraphs of his 1935 paper:~\cite{Einst35}
\begin{quotation}
Thus, in the book just mentioned, essential use is made of
the concept of force, which in the relativity theory has no such
direct significance as it has in classical mechanics. This is connected
with the fact that, in the latter, the force is to be considered
as a given function of the coordinates of all the particles,
which is obviously not possible in the relativity theory. Therefore
I have avoided introducing the force concept.

Furthermore, I was concerned with avoiding making any
assumption concerning the transformation character of impulse
and energy with respect to a Lorentz transformation.
\end{quotation}



\begin{thebibliography}{99}
\bibitem{note}Equation (\ref{eq:consGamma}) implies the conservation of $\sum(\gamma(u)+\text{const})$. The value of the constant is chosen so that the expression vanishes for $u\to 0$.
\end{thebibliography}
\end{document}